\documentclass[12pt]{article}

\usepackage{amsmath,amsthm,amssymb,amscd,a4wide}
\usepackage[dvips]{graphicx}

\usepackage{mybbold}    

\newcommand{\ud}{\text{d}}
\newcommand{\ui}{\text{i}}
\newcommand{\ue}{\text{e}}

\newcommand{\SU}{\mathrm{SU}}

\newcommand{\su}{\mathrm{su}}
\newcommand{\SO}{\mathrm{SO}}
\newcommand{\SP}{\mathrm{Sp}}
\newcommand{\Ad}{\mathrm{Ad}}
\newcommand{\Stwo}{{\mathrm{S}^2}}
\newcommand{\id}{\operatorname{id}}

\newcommand{\de}{\delta}
\newcommand{\De}{\Delta}

\newcommand{\la}{\lambda}
\newcommand{\om}{\omega}
\newcommand{\Om}{\Omega}
\newcommand{\si}{\sigma}

\newcommand{\ve}{\varepsilon}
\newcommand{\vp}{\varphi}

\newcommand{\cH}{{\mathcal H}}

\newcommand{\cS}{{\mathcal S}}

\newcommand{\vecA}{\boldsymbol{A}} 
\newcommand{\vecB}{\boldsymbol{B}}
\newcommand{\vecb}{\boldsymbol{b}}
\newcommand{\vecC}{\boldsymbol{C}}

\newcommand{\vecL}{\boldsymbol{L}}
\newcommand{\vecj}{\boldsymbol{j}}
\newcommand{\vecJ}{\boldsymbol{J}}
\newcommand{\vecm}{\boldsymbol{m}}
\newcommand{\vecn}{\boldsymbol{n}}
\newcommand{\vecs}{\boldsymbol{s}}
\newcommand{\vecS}{\boldsymbol{S}}

\newcommand{\vecx}{\boldsymbol{x}}
\newcommand{\vecX}{\boldsymbol{X}}
\newcommand{\vecy}{\boldsymbol{y}}
\newcommand{\vecz}{\boldsymbol{z}}
\newcommand{\vecp}{\boldsymbol{p}}
\newcommand{\vecP}{\boldsymbol{P}}
\newcommand{\vecq}{\boldsymbol{q}}
 
\newcommand{\vecsig}{\boldsymbol{\sigma}}

\newcommand{\vecxi}{\boldsymbol{\xi}}

\newcommand{\rz}{{\mathbb R}}
\newcommand{\nz}{{\mathbb N}}
\newcommand{\kz}{{\mathbb C}}

\newcommand{\kommentar}[1]{}

\DeclareMathOperator{\tr}{Tr}
\DeclareMathOperator{\mtr}{tr}
\DeclareMathOperator{\vol}{vol}

\newcommand{\eins}{\mathmybb{1}}

\numberwithin{equation}{section}

\theoremstyle{definition}

\begin{document}

\thispagestyle{empty}

\noindent
ULM-TP/01-1\\
January 2001\\

\vspace*{1cm}

\begin{center}

{\LARGE\bf Quantum and classical ergodicity  \\
\vspace*{3mm}
of spinning particles} \\
\vspace*{3cm}
{\large Jens Bolte}%
\footnote{E-mail address: {\tt bol@physik.uni-ulm.de}},
{\large Rainer Glaser}%
\footnote{E-mail address: {\tt gla@physik.uni-ulm.de}} {\large and}
{\large Stefan Keppeler}%
\footnote{E-mail address: {\tt kep@physik.uni-ulm.de}}

\vspace*{1cm}

Abteilung Theoretische Physik\\
Universit\"at Ulm, Albert-Einstein-Allee 11\\
D-89069 Ulm, Germany 
\end{center}

\vfill

\begin{abstract}
We give a formulation of quantum ergodicity for Pauli Hamiltonians with 
arbitrary spin in terms of a Wigner-Weyl calculus. The corresponding 
classical phase space is the direct product of the phase space of the 
translational degrees of freedom and the two-sphere. On this product space 
we introduce a combination of the translational motion and classical spin 
precession. We prove quantum ergodicity under the condition that this 
product flow is ergodic.
\end{abstract}

\newpage

\section{Introduction}
\label{1sec}
Quantum ergodicity describes the equidistribution of eigenfunctions 
in the semiclassical limit ($\hbar\to 0$), in the sense that almost all 
expectation values of quantum observables tend to a classical mean value of 
the corresponding classical observable. 
In particular, the Wigner functions associated with almost all eigenfunctions 
of the Hamiltonian become equidistributed on the energy shell in phase space.
A sufficient condition for quantum ergodicity to hold 
is ergodicity of the corresponding classical dynamics, as 
was first stated by Shnirelman \cite{Shn74}. It is remarkable that quantum 
ergodicity is one of the few results in quantum chaos for which there exist 
mathematical proofs; the first complete ones were given by 
Zelditch \cite{Zel87} and Colin de Verdi\`ere \cite{CdV85}. These 
are results for scalar Hamiltonians, i.e. for the Laplacian on compact 
manifolds or, more generally, for Schr\"odinger operators in $\rz^d$ 
\cite{HelMarRob87}. The case of quantum billiards was considered in 
\cite{GerLei93,ZelZwo96}.

If one wants to describe particles with internal degrees of freedom, such as
spin, one has to deal with matrix valued Hamiltonians. Since spin is a 
purely quantum mechanical property, it is a priori not obvious what should 
serve as the corresponding classical system. In \cite{BolGla00} quantum 
ergodicity for Pauli Hamiltonians with spin 1/2 was proven under the condition
that a combination of the classical translational dynamics and a quantum
spin dynamics driven by the translational motion is ergodic. This kind of 
mixed classical/quantum description in terms of a skew product dynamics
arises naturally in the semiclassical 
analysis of Hamiltonians with spin \cite{BolKep98,BolKep99a,BolKep99b}. 

Although spin is a genuinely quantum mechanical property, there exists 
an intuitive classical analogue given by the so-called vector model which 
provides a picture of spin as a vector of fixed length whose dynamics 
is similar to that of angular momentum. The mathematical description of this 
model goes back to Thomas \cite{Tho27}, who gave an equation of motion for a 
classical particle with a magnetic moment precessing in external fields. 
In a semiclassical analysis of spinning particles various objects can be
related to this classical spin precession \cite{RubKel63,BolKep99a,Spo00}.
In the vector model the equations of motion are of first order in time, 
and for pure spin precession can be formulated in a Hamiltonian framework.
Thus, after suitable normalisation of the spin vector, the unit sphere 
$\Stwo$ serves as the phase space of classical spin precession.


Based on ideas of Stratonovich \cite{Str57}, Gracia-Bond{\'\i}a and V\'arilly 
\cite{GraVar88,VarGra89} developed a Wigner-Weyl calculus for arbitrary 
spin $j$, in which quantum mechanical observables, i.e. hermitian 
$(2j+1)\times(2j+1)$ matrices, are represented in terms of functions on 
the phase space $\Stwo$. We use this description to rephrase the problem of 
quantum ergodicity for non-relativistic particles with spin 
and to prove for arbitrary spin the respective statement 
under the condition that a different skew product of translational and spin
dynamics, where now the spin part is given by classical spin precession, 
is ergodic.

The paper is organised as follows. In section \ref{2sec} we briefly review 
properties of the Weyl correspondence for systems with only either 
translational or spin degrees of freedom. In section \ref{3sec} we then
discuss how in the semiclassical limit of systems with both kinds of 
degrees of freedom the motion is governed by the skew product dynamics. 
We also state a corresponding Egorov theorem. Section \ref{4sec} is 
devoted to the proof of quantum ergodicity for Pauli Hamiltonians with 
arbitrary spin. Some particular representations of the Wigner-Weyl transform 
for spinors and a relation between ergodic properties of the two types of skew 
products are discussed in two appendices.

\section{Weyl quantisation and classical limit}
\label{2sec}
The Weyl quantisation of systems in $\rz^d$ without spin, i.e.\ taking
only translational degrees of freedom into account, is based on unitary
irreducible representations of the Weyl-Heisenberg operators
\begin{equation}
\label{Weylop}
\rho(\vecp,\vecq) := \ue^{\frac{\ui}{\hbar}(\vecq\cdot\hat{\vecP}+
\vecp\cdot\hat{\vecX})}\ ,
\end{equation}
where $\hat P_k,\hat X_k$, $k=1,\dots,d$, are momentum and position operators, 
respectively, satisfying canonical commutation relations. According to the 
Stone-von Neumann theorem such representations are unitarily equivalent
to the Schr\"odinger representation $\rho_S(\vecp,\vecq)$ of (\ref{Weylop})
on the Hilbert space $\cH_{\rm trans}=L^2(\rz^d)$, in which $\hat X_k$ is 
realised as a multiplication operator and $\hat P_k=\frac{\hbar}{\ui}
\frac{\partial}{\partial x_k}$. The Schr\"odinger representation hence acts 
on $\psi\in L^2(\rz^d)$ as
\begin{equation}
\label{Schroerep}
\bigl(\rho_S(\vecp,\vecq)\,\psi\bigr)(\vecx) = \ue^{\frac{\ui}{\hbar}
\vecp\cdot(\vecx+\frac{1}{2}\vecq)}\,\psi(\vecx+\vecq)\ .
\end{equation}
See \cite{Fol89} for further details.

A classical observable $B(\vecp,\vecx)$ is a real valued function on
phase space $\rz^d\times\rz^d$, to which the Weyl quantisation assigns a
symmetric operator $\hat B$ on $L^2(\rz^d)$,
\begin{equation}
\label{W-quant}
\hat B := \frac{1}{(2\pi\hbar)^d}\int_{\rz^d}\int_{\rz^d}\tilde B (\vecp,
\vecq)\,\rho_S(\vecp,\vecq)\ \ud p\,\ud q \ ,
\end{equation}
where
\begin{equation}
\label{Fouriert}
\tilde B (\vecp,\vecq) := \frac{1}{(2\pi\hbar)^d}\int_{\rz^d}\int_{\rz^d}
B(\vecxi,\vecz)\,\ue^{-\frac{\ui}{\hbar}(\vecxi\cdot\vecq+\vecz\cdot\vecp)}
\ \ud\xi\,\ud z
\end{equation}
denotes the Fourier transform of the classical observable. The operation 
of $\hat B$ on (suitable) $\psi\in L^2(\rz^d)$ then reads
\begin{equation}
\label{W-op}
\hat B\psi(\vecx) = \frac{1}{(2\pi\hbar)^d}\int_{\rz^d}\int_{\rz^d}
\ue^{\frac{\ui}{\hbar}\vecp\cdot(\vecx-\vecy)}\,B\bigl(\vecp,\tfrac{1}{2}
(\vecx+\vecy)\bigr)\,\psi(\vecy)\ \ud y\,\ud p\ ,
\end{equation}
In fact, $B$ can even be a (tempered) distribution on phase space, 
$B\in\cS'(\rz^d\times\rz^d)$; in this case the operator $\hat B$ is 
defined on the domain $\cS(\rz^d)\subset L^2(\rz^d)$. Reversing the above 
reasoning, one notices that the Schwartz kernel theorem allows to represent
every continuous linear map $\hat B$ from $\cS(\rz^d)$ into $\cS'(\rz^d)$
as a Weyl operator (\ref{W-op}) with symbol $B(\vecp,\vecx)$, which
in general is a distribution on phase space (see, e.g., \cite{Fol89}). The 
map $\hat B\mapsto B(\vecp,\vecx)$ can be made more explicit if $\hat B$ is 
a trace class operator on $L^2(\rz^d)$. Since $\rho_S (\vecp,\vecq)$ is 
bounded, $\hat B\,\rho_S(\vecp,\vecq)$ is also of trace class, and  
\begin{equation}
\label{symbolcalc}
\tilde B(\vecp,\vecq) = \tr \bigl(\hat B\,\rho_S(\vecp,\vecq) \bigr) \ ,
\end{equation}
where $\tr(\cdot)$ denotes the operator trace on $\cH_{\rm trans}$. We remark 
that the symbol of a Weyl operator can in principle depend on $\hbar$. In 
such a case a direct interpretation of $B(\vecp,\vecx)$ as a classical 
observable is inappropriate. However, if one restricts the class of 
observables suitably, the symbols possess asymptotic expansions as 
$\hbar\to 0$,  
\begin{equation}
\label{symbasymp}
B(\vecp,\vecx;\hbar) \sim \sum_{k\geq 0}\hbar^k\,B_k(\vecp,\vecx) \ ;
\end{equation}
for the precise meaning of (\ref{symbasymp}) see, e.g., \cite{Rob87,DimSjo99}.
Now the principal symbol $B_0(\vecp,\vecx)$ is a function on phase
space independent of $\hbar$ that serves as the corresponding classical 
observable.

Summarising its most important properties, the Weyl quantisation can be
characterised by the following properties (see, e.g., 
\cite{Fol89,Rob87,DimSjo99}):
\begin{enumerate}
\item[(i)] The map $\hat B\mapsto B(\vecp,\vecx)$ is linear.
\item[(ii)] $\hat B^\dagger\mapsto\overline{B(\vecp,\vecx)}$.
\item[(iii)] If $\hat B$ is of trace class, then
\begin{equation}
\label{W-trace}
\tr\hat B = \frac{1}{(2\pi\hbar)^d}\int_{\rz^d}\int_{\rz^d}B(\vecp,\vecx)
\ \ud p\,\ud x\ .
\end{equation}
\item[(iv)] Covariance: If $G\in\SP(2d,\rz)$ is a linear canonical 
transformation with metaplectic representation $\pi_M(G)$, then $\hat B\mapsto
B(\vecp,\vecx)$ implies
\begin{equation}
\label{W-cov}
\pi_M(G)^\dagger\,\hat B\,\pi_M(G)\mapsto B(G(\vecp,\vecx))\ .
\end{equation}
\end{enumerate}

We now turn attention to a quantum mechanical spin, without translational
degrees of freedom. Its (pure) states are described by vectors in the 
Hilbert space $\cH_{\rm spin}=\kz^{2j+1}$, where $j\in\tfrac{1}{2}\nz$ denotes
the (fixed) spin quantum number. Spin observables are therefore hermitian
$(2j+1)\times (2j+1)$ matrices. The corresponding classical phase space
is given by the two-sphere ${\mathrm S}^{2}$, considered as a symplectic 
manifold whose symplectic structure is provided by the area two-form. In 
spherical coordinates $(\theta,\vp)$ for ${\mathrm S}^{2}$ a symplectic 
chart is given by $(p,q)=(-\cos\theta,\vp)\in (-1,1)\times (0,2\pi)$. For 
reasons of convenient normalisation we here choose the symplectic two-form 
as 
\begin{equation}
\label{sympform}
\om(p,q) := \frac{1}{\hbar\sqrt{j(j+1)}}\,\ud p\wedge\ud q =
\frac{1}{\hbar\sqrt{j(j+1)}}\,\sin\theta\,\ud\theta\wedge\ud\vp\ .
\end{equation}
In the following we will often denote points of ${\mathrm S}^{2}$ by
$\vecn\in\rz^3$ with $|\vecn|=1$ so that in spherical coordinates $\vecn =
(\sin\theta\cos\vp,\sin\theta\sin\vp,\cos\theta)^T$. In these coordinates
the normalised area two-form is given by $\ud\mu_{{\mathrm S}^{2}}(\vecn) =
\frac{1}{4\pi}\sin\theta\,\ud\theta\wedge\ud\vp$. Below we will show that
the quantum mechanical spin operator $\hat{\vecS}$ is then associated with 
the vector $\vecs := \hbar\sqrt{j(j+1)}\,\vecn \in\rz^3$ of length 
$\hbar\sqrt{j(j+1)}$; in other words, $\vecn=\vecs/|\vecs|\in{\mathrm S}^{2}$ 
represents a normalised `classical' spin vector.

A Weyl correspondence for spin, with properties as close as possible to 
the ones discussed above in the case of translational degrees of freedom,
has to assign a real valued function on ${\mathrm S}^{2}$ to a hermitian 
$(2j+1)\times (2j+1)$ matrix in such a way that properties analogous to
(i)--(iv) of above hold. Based on ideas of Stratonovich \cite{Str57}, 
V\'arilly and Gracia-Bond{\'\i}a \cite{GraVar88,VarGra89} 
have constructed such a Weyl
correspondence. In their approach quantum mechanical observables $\hat b$
are represented as 
\begin{equation}
\label{spinWeyl}
\hat b = (2j+1)\int_{{\mathrm S}^{2}} b(\vecn)\,\hat\De_j(\vecn)\ 
\ud\mu_{{\mathrm S}^{2}}(\vecn)\ .
\end{equation}
Here $\hat\De_j(\vecn)$ is a function on ${\mathrm S}^{2}$ taking values in 
the hermitian $(2j+1)\times (2j+1)$ matrices, and the symbol $b(\vecn)\in 
L^2({\mathrm S}^{2})$ can be obtained from the observable $\hat b$ in 
analogy to (\ref{symbolcalc}),
\begin{equation}
\label{spinsymbol}
b(\vecn) = \mtr \bigl(\hat b\,\hat\De_j(\vecn)\bigr) \ ,
\end{equation}
with $\mtr(\cdot)$ meaning the matrix trace. In \cite{VarGra89} it is shown 
that there are $2^{2j}$ possibilities for a Weyl correspondence with a kernel 
$\hat\De_j(\vecn)$ that fulfills
\begin{itemize}
\item[(a)] $\hat\De_j(\vecn)^\dagger =\hat\De_j(\vecn)$ for all 
$\vecn\in{\mathrm S}^{2}$.
\item[(b)] $(2j+1)\int_{{\mathrm S}^{2}}\hat\De_j(\vecn)\ 
\ud\mu_{{\mathrm S}^{2}}(\vecn)=\eins_{2j+1}$.
\item[(c)] Covariance: If $g\in\SU(2)$ and $\pi_j$ denotes the 
$(2j+1)$-dimensional unitary irreducible representation of $\SU(2)$, then 
$\hat\De_j(\vp(g)\vecn)=\pi_j(g)\,\hat\De_j(\vecn)\,\pi_j(g)^\dagger$.
\end{itemize}
In the covariance property of the kernel $\vp$ denotes the covering map
from $\SU(2)$ to $\SO(3)$: For every $g\in\SU(2)$ one defines the adjoint map 
$\Ad_g$ as the linear map of the Lie algebra $\su(2)$ given by $X\mapsto 
g^\dagger\,X\,g$. Expanding $X\in\su(2)$ in terms of the Pauli matrices, 
$X=\vecx\cdot\vecsig$, $\vecx\in\rz^3$, the adjoint map operates as 
$\Ad_g(\vecx\cdot\vecsig)=(\vp(g)\vecx)\cdot\vecsig$. In this way every 
$g\in\SU(2)$ is mapped to some $\vp(g)\in\SO(3)$; in fact this map is 
two-to-one and provides the universal covering of $\SO(3)$ by $\SU(2)$. 
Hence $\vp(g)\vecn$ means the rotation of $\vecn\in\rz^3$ (with $|\vecn|=1$) 
by $\vp(g)\in\SO(3)$. In \cite{VarGra89} one choice out of the $2^{2j}$ 
possibilities to define $\hat\De_j(\vecn)$ is singled out due to its 
connection with spin-coherent states; however, for our purposes this 
particular choice is not important and, thus, will not be specified here. 
For illustration, some explicit examples are shown in appendix \ref{kernels}. 

The properties (a)--(c) of $\hat\De_j(\vecn)$ listed above now imply properties
of the symbols $b(\vecn)$ that are closely analogous to those of symbols
in the case of translational degrees of freedom: \label{symbProp}
\begin{itemize}
\item[(i)] 
  The map $\hat b\mapsto b(\vecn)$ is linear.
\item[(ii)] 
  $\hat b^\dagger\mapsto\overline{b(\vecn)}$.
\item[(iii)]  
  $\mtr\hat b=(2j+1)\int_{{\mathrm S}^{2}}b(\vecn)\,
        \ud\mu_{{\mathrm S}^{2}}(\vecn)$.
\item[(iv)]  
  For every $g\in\SU(2)$: 
  $\pi_j(g)^\dagger\,\hat b\,\pi_j(g)\mapsto b(\vp(g)\vecn)$.
\end{itemize}
In addition one obtains the simple relation 
\begin{equation}
\label{products}
  \mtr (\hat{a}\hat{b}) 
  = (2j+1) \int_{\Stwo} a(\vecn)\,b(\vecn) \ \ud\mu_{\Stwo}(\vecn) 
\end{equation}
for the trace of a matrix product.

The covariance of the quantisation as expressed by (iv) relates the quantum
mechanical time evolution of a spin to a classical dynamics (see also
\cite{VarGra89}): Let $\ud\pi_j$ denote the $(2j+1)$-dimensional derived 
representation of $\su(2)$, i.e. $\ud\pi_j(X)=\frac{1}{\ui}\frac{\ud}{\ud\la}
\pi_j(\ue^{\ui\la X})|_{\la =0}$. Then the spin operators $\hat S_k$, 
$k=1,2,3$, are given by $\hat S_k=\frac{\hbar}{2}\ud\pi_j(\si_k)$, with
commutation relations $[\hat S_k,\hat S_l]=\ui\hbar\ve_{klm}\hat S_m$. The 
Weyl correspondence now assigns the vector valued symbol $\vecs =
\hbar\sqrt{j(j+1)}\,\vecn$ to the vector $\hat{\vecS}$ of these quantum 
mechanical spin operators, see appendix \ref{kernels}.

A typical (time independent) quantum Hamiltonian is given by
\begin{equation}
\label{spinHam}
\hat H_{\rm spin} = \hat{\vecS}\cdot\vecC = \frac{\hbar}{2}\,\vecC\cdot
\ud\pi_j(\vecsig) \ ,
\end{equation}
where $\vecC\in\rz^3$ is some constant vector. This Hamiltonian generates
a quantum mechanical time evolution that is governed by the unitary operators
\begin{equation}
\label{spindyn}
\ue^{-\frac{\ui}{\hbar}\hat H_{\rm spin}t} = \pi_j\bigl(\ue^{-\frac{\ui}{2}
\vecC\cdot\vecsig t}\bigr) \ ,
\end{equation}
so that a time evolved observable reads
\begin{equation}
\label{qmevolve}
\hat b(t) = \ue^{\frac{\ui}{\hbar}\hat H_{\rm spin}t}\,\hat b\,
\ue^{-\frac{\ui}{\hbar}\hat H_{\rm spin}t} = \pi_j\bigl(\ue^{-\frac{\ui}{2}\vecC
\cdot\vecsig t}\bigr)^\dagger\,\hat b\,\pi_j\bigl(\ue^{-\frac{\ui}{2}\vecC
\cdot\vecsig t}\bigr) \ .
\end{equation}
If $b(\vecn)$ is the classical observable associated with $\hat b$, the 
relation (\ref{qmevolve}) and the covariance property (iv) hence 
assign the classical observable
\begin{equation}
\label{spinEgorov}
b(t)(\vecn) = b\bigl(\vp(\ue^{-\frac{\ui}{2}\vecC\cdot\vecsig t})\vecn\bigr)
\end{equation}
to $\hat b(t)$. In this way one is provided with a dynamics 
$\vecn\mapsto\vecn(t)=\vp(\ue^{-\frac{\ui}{2}\vecC\cdot\vecsig t})\vecn$ on 
${\mathrm S}^{2}$ that is governed by the equations of motion
\begin{equation}
\label{clprecess}
\dot{\vecn}(t) = \vecC\times\vecn(t)
\end{equation}
describing a precession of $\vecn(t)$ about $\vecC$. On the other hand, the 
Weyl correspondence associates the classical Hamiltonian $H_{\rm spin}(\vecn)=
\vecs\cdot\vecC$ to the quantum Hamiltonian (\ref{spinHam}). Together with 
the Poisson bracket $\{\cdot,\cdot\}_{{\mathrm S}^{2}}$ derived from the 
symplectic structure on ${\mathrm S}^{2}$ given by $\om$ one therefore
identifies (\ref{clprecess}) as the equations of motion generated by
$H_{\rm spin}$, i.e.
\begin{equation}
\label{Hspincldyn}
\dot{\vecn}(t) = \{H_{\rm spin},\vecn(t)\}_{{\mathrm S}^{2}} 
= \vecC\times\vecn(t) \ .
\end{equation}
Thus the assignment (\ref{spinsymbol}) of a function on ${\mathrm S}^{2}$ 
to a quantum observable commutes with the (quantum mechanical or classical,
respectively) time evolution, see also appendix \ref{ergodicities}. This 
property is a special feature of the Weyl correspondence for spin. In the 
case of translational degrees of freedom the analogous relation, known as 
the Egorov theorem \cite{Ego69}, only holds for the assignment of the 
principal symbol $B_0(\vecp,\vecx)$, and not of the full symbol 
$B(\vecp,\vecx)$, to a quantum observable.
   
\section{Coupling of spin and translational motion}
\label{3sec}
The quantum mechanical description of particles with spin requires to
couple translational and spin degrees of freedom. The relevant Hilbert
space $\cH$ therefore has to be the tensor product of $\cH_{\rm trans}$ and 
$\cH_{\rm spin}$, i.e.\ $\cH= L^2(\rz^d)\otimes\kz^{2j+1}$. The state vectors
of a spinning particle are hence $(2j+1)$-component spinors whose components 
are square integrable over $\rz^d$. Observables are represented by 
self-adjoint operators on $\cH$, which in Weyl representation read
\begin{equation}
\label{BWeyl}
\hat B\psi(\vecx) = \frac{1}{(2\pi\hbar)^d}\int_{\rz^d}\int_{\rz^d}
\ue^{\frac{\ui}{\hbar}\vecp\cdot(\vecx-\vecy)}\,B\bigl(\vecp,\tfrac{1}{2}
(\vecx+\vecy)\bigr)\,\psi(\vecy)\ \ud y\,\ud p\ .
\end{equation}
Here the symbol $B(\vecp,\vecx)$ is in general a $(2j+1)\times (2j+1)$ matrix 
valued distribution on phase space $\rz^d\times\rz^d$. The operators 
(\ref{BWeyl}) then act on $\psi\in\cS(\rz^d)\otimes\kz^{2j+1}$. 

In a non-relativistic context the quantum dynamics is generated by the 
(Pauli) Hamiltonian
\begin{equation}
\label{PauliH}
\hat H_P = \hat H_{\rm trans}\eins_{2j+1} + \frac{\hbar}{2}\,\ud\pi_j(\vecsig)
\cdot\hat{\vecC}
\end{equation}
that consists of a scalar part $\hat H_{\rm trans}$ describing the dynamics of
the translational degrees of freedom, and a genuinely matrix valued part
that contains a coupling of spin to the translational motion; here 
$\hat C_k$ are Weyl quantisations of suitable functions $C_k(\vecp,\vecx)$
on phase space. Typically, $\hat H_{\rm trans}$ is of the form
\begin{equation}
\label{Htrans}
\hat H_{\rm trans} = \frac{1}{2m}\Bigl(\frac{\hbar}{\ui}\nabla -\frac{e}{c}
\vecA(\vecx)\Bigr)^2 +e\phi(\vecx)\ ,
\end{equation}
with (static) electromagnetic potentials $\vecA,\phi$, i.e.\ it is a Weyl 
quantisation of the classical Hamiltonian
\begin{equation}
\label{H0}
H_0(\vecp,\vecx) = \frac{1}{2m}\Bigl(\vecp -\frac{e}{c}\vecA(\vecx)
\Bigr)^2 +e\phi(\vecx)\ .
\end{equation}
Examples for the coupling of spin to the translational motion are given 
by a coupling to an external magnetic field, $\vecC(\vecp,\vecx)=
-\tfrac{e}{2mc}\vecB(\vecx)$, or a spin-orbit coupling $\vecC(\vecp,\vecx)=
\frac{1}{4m^2 c^2|\vecx|}\frac{\ud\phi(|\vecx|)}{\ud |\vecx|}
(\vecx\times\vecp)$. For the following the specific form of the Hamiltonian, 
however, is not relevant. We only require $H_0(\vecp,\vecx)$ to fulfill 
certain criteria that guarantee $\hat H_P$ to be (essentially) self-adjoint, 
bounded from below, and to possess a purely discrete spectrum in some 
interval $[E-\ve,E+\ve]$. For details see \cite{BolGla00}.

For simplicity the further observables that we are going to consider shall
be bounded Weyl operators on $\cH$. Their symbols are smooth, matrix valued
functions on phase space, which may also depend on $\hbar$ in such a way 
that an asymptotic expansion ($\hbar\to 0$)
\begin{equation}
\label{symbasymp1}
B(\vecp,\vecx;\hbar) \sim \sum_{k\geq 0}\hbar^k\,B_k(\vecp,\vecx) 
\end{equation}
analogous to (\ref{symbasymp}) holds (see \cite{BolGla00} for further 
details). As far as the translational degrees of freedom are concerned, the 
classical observable corresponding to $\hat B$ is then given by its principal 
symbol $B_0(\vecp,\vecx)$, which is a function on phase space taking values 
in the hermitian $(2j+1)\times (2j+1)$ matrices. Thus spin is still described
on a quantum mechanical level. As an example of an observable, although an 
unbounded one, that illustrates this situation consider the total angular 
momentum $\hat{\vecJ} = \hat{\vecL} + \hat{\vecS}$, whose symbol
\begin{equation}
\label{Jsymb}
\vecJ(\vecp,\vecx) = (\vecx\times\vecp)\,\eins_{2j+1} + \hbar\,\frac{1}{2}\,
\ud\pi_j(\vecsig)
\end{equation}
has a principal part that consists of orbital angular momentum, and a
sub-principal part given by spin.

The relation between the quantum and classical time evolution, i.e.\ the 
relevant Egorov theorem, has in this context been derived in \cite{BolGla00}. 
In order to state this version of the Egorov theorem, let $d(\vecp,\vecx,t)
\in\SU(2)$ be the solution of the spin-transport equation \cite{BolKep99a}
\begin{equation}
\label{spintrans}
\dot{d}(\vecp,\vecx,t)+\frac{\ui}{2}\vecC(\Phi^t(\vecp,\vecx))\cdot\vecsig\,
d(\vecp,\vecx,t)=0\qquad\text{with}\qquad d(\vecp,\vecx,0)=\eins_2\ ,
\end{equation}
where $\Phi^t(\vecp,\vecx)=(\vecp(t),\vecx(t))$ denotes the classical flow
on the phase space $\rz^d\times\rz^d$ generated by the classical 
translational Hamiltonian $H_0(\vecp,\vecx)$, i.e. $(\vecp(t),\vecx(t))$ are 
solutions of Hamilton's equations of motion with initial condition 
$(\vecp(0),\vecx(0))=(\vecp,\vecx)$. Hence, a comparison with (\ref{spindyn}) 
reveals that $\pi_j(d(\vecp,\vecx,t))$ is the quantum mechanical propagator 
for the spin degrees of freedom along the trajectories $\Phi^t(\vecp,\vecx)$ 
of the classical translational motion. The Egorov theorem relevant in the 
present situation now states that the time evolution $\hat B(t)$, generated 
by a Pauli Hamiltonian (\ref{PauliH}), of an observable (\ref{BWeyl}) with 
a symbol allowing for an asymptotic expansion (\ref{symbasymp1}), is again 
an operator of the same type and its principal symbol $B_0(t)(\vecp,\vecx)$ 
reads
\begin{equation}
\label{PauliEgorov}
B_0(t)(\vecp,\vecx) = \pi_j\bigl(d(\vecp,\vecx,t)\bigr)^\dagger\,B_0(\Phi^t
(\vecp,\vecx))\,\pi_j\bigl(d(\vecp,\vecx,t)\bigr)\ .
\end{equation}
In this expression the translational degrees of freedom are obviously 
propagated classically by means of the flow $\Phi^t$. In contrast, a
glance at the relation (\ref{qmevolve}) reveals that the dynamics of the 
spin degrees of freedom are genuinely quantum mechanical, but are driven by 
the classical translational flow and take place along the trajectories of 
this flow. This observation reflects the mixed (classical/quantum) level
of description of the translational and spin degrees of freedom.
 
We now proceed to describe spin within the framework of the Weyl 
correspondence outlined in the previous section. Thus, according to
(\ref{BWeyl}) and (\ref{spinWeyl}) quantum mechanical observables $\hat B$
are represented as 
\begin{equation}
\label{BWeylmod}
\hat B\psi(\vecx) = \frac{2j+1}{(2\pi\hbar)^d}\int_{\rz^d}\int_{\rz^d}
\int_{{\mathrm S}^2}\ue^{\frac{\ui}{\hbar}\vecp\cdot(\vecx-\vecy)}\,
b\bigl(\vecp,\tfrac{1}{2}(\vecx+\vecy),\vecn\bigr)\,\hat\De_j(\vecn)\,
\psi(\vecy)\ \ud y\,\ud p\,\ud\mu_{{\mathrm S}^2}(\vecn)\ ,
\end{equation}
where according to (\ref{spinsymbol})
\begin{equation}
\label{transspinsymbol}
b(\vecp,\vecx,\vecn) = \mtr \bigl(B(\vecp,\vecx)\,\hat\De_j(\vecn)\bigr)
\end{equation}
is the associated scalar symbol; e.g, in the case of the operator 
$\hat{\vecJ}$ of total angular momentum the scalar symbol related to the
matrix valued one (\ref{Jsymb}) is given by
\begin{equation} 
\label{Jscalarsymb} 
\vecj(\vecp,\vecx,\vecn) = \vecx\times\vecp + \hbar\,\sqrt{j(j+1)}\,\vecn\ .
\end{equation}

When applied to the principal symbol $B_0(\vecp,\vecx)$, the relation 
(\ref{transspinsymbol}) now associates with every quantum observable 
$\hat B$ of the type under consideration a classical observable 
$b_0(\vecp,\vecx,\vecn)$, which is a function on the combined phase 
space $\rz^d\times\rz^d\times {\mathrm S}^2$. In particular, the Egorov 
theorem can be brought into the following form: According to 
(\ref{PauliEgorov}) the matrix valued principal symbol of $\hat B(t)$ is 
of the same type as the right-hand side of (\ref{qmevolve}). Hence, due to 
the covariance property of the spin-Weyl correspondence with respect to 
$\SU(2)$, the associated scalar principal symbol $b_0(t)(\vecp,\vecx,\vecn) 
= \mtr(B_0(t)(\vecp,\vecx) \hat{\Delta}_j(\vecn))$ reads
\begin{equation}
\label{scalarEgorov}
b_0(t)(\vecp,\vecx,\vecn) = b_0 \bigl(\Phi^t(\vecp,\vecx),\vp(d(\vecp,\vecx,t))
\vecn\bigr) = b_0(\vecp(t),\vecx(t),\vecn(t))\ .
\end{equation}
The classical observable corresponding to the quantum observable $\hat B(t)$ 
at time $t$ therefore follows from the respective classical observable
at time zero by a purely classical time evolution on the combined phase space 
of translational and spin degrees of freedom. The combined flow on 
$\rz^d\times\rz^d\times {\mathrm S}^2$, denoted as $Y_{\rm cl}^t$, is 
therefore given by the skew product,
\begin{equation}
\label{skew2}
Y_{\rm cl}^t(\vecp,\vecx,\vecn) 
:= \bigl(\Phi^t(\vecp,\vecx),\vp(d(\vecp,\vecx,t)) \vecn\bigr) 
= (\vecp(t),\vecx(t),\vecn(t)) \, ,
\end{equation}
%
and the scalar Egorov relation (\ref{scalarEgorov}) reads
\begin{equation}
\label{scalarEgorov2}
b_0(t)(\vecp,\vecx,\vecn) = b_0\bigl(Y_{\rm cl}^t(\vecp,\vecx,\vecn)\bigr)\, .
\end{equation}
See, e.g., \cite{CorFomSin82} for a general discussion of skew products.
The spin part $\vecn\mapsto\vecn(t)=\vp(d(\vecp,\vecx,t))\vecn$ of the 
combined classical dynamics takes place along the trajectories $\Phi^t
(\vecp,\vecx)$ of the translational motion and describes a precession of
$\vecn(t)$ about the instantaneous axis $\vecC(\Phi^t(\vecp,\vecx))$, compare
(\ref{clprecess}). This is a non-relativistic version of the Thomas 
precession of spin \cite{Tho27}.

\section{Quantum ergodicity}
\label{4sec}
Quantum ergodicity for Pauli Hamiltonians of the form (\ref{PauliH})
with spin $1/2$ was derived in \cite{BolGla00} under the condition that the 
underlying mixed classical/quantum system be ergodic. For this purpose, the 
two dynamics inherent in (\ref{PauliEgorov}) had to be combined into a single 
skew product flow $Y^t$ on $\rz^d \times \rz^d \times \SU(2)$ via
\begin{equation}
\label{skew1}
  Y^t((\vecp,\vecx),g):=(\Phi^t(\vecp,\vecx),d(\vecp,\vecx,t)g) \ ,\qquad 
  (\vecp,\vecx,g) \in \rz^d\times\rz^d\times\SU(2)\ ,
\end{equation}
see also \cite{BolKep99b} for further details. Introducing Liouville measure 
(the microcanonical distribution)
\begin{equation}
\label{Liouville}
\ud\mu_E(\vecp,\vecx) := \frac{1}{\vol\Om_E}\,\de\bigl(H_0(\vecp,\vecx)-E
\bigr)\ \ud p\,\ud x
\end{equation}
on the energy shell
\begin{equation}
\label{Yergod}
  \Om_E := \left\{ (\vecp,\vecx) \in \rz^d \times \rz^d \, 
                      | \ H_0(\vecp,\vecx) = E \right\} \ ,
\end{equation}
one observes that $Y^t$ leaves the product measure $\mu := \mu_E\times\mu_H$ 
invariant, where $\mu_H$ denotes the normalised Haar measure on $\SU(2)$.
Now ergodicity of $Y^t$ on $\Om_E \times \SU(2)$ with respect to $\mu$ means 
that for every integrable function $F\in L^1(\Om_E\times\SU(2),\ud\mu)$ the 
relation 
\begin{equation} 
\label{su2ergodic}
 \lim_{T \to \infty} \frac{1}{T} \int_0^T F(Y^t(\vecxi,\vecy,h)) \ \ud t = 
 \int_{\Om_E \times \SU(2)} F(\vecp,\vecx,g) \ \ud \mu(\vecp,\vecx,g) 
\end{equation}
holds for $\mu$-almost all initial conditions $(\vecxi,\vecy,h)\in\Om_E\times
\SU(2)$. Provided that ergodicity of $Y^t$ and certain properties of the 
principal symbol $H_0$ of the quantum Hamiltonian hold, 
it was shown in \cite{BolGla00} that in every sequence $\{\psi_k \, | \, E_k 
\in I(E,\hbar) \}$ of orthonormal eigenspinors of the Hamiltonian 
$\hat H_P$ with eigenvalues $E_k$ in an interval $I(E,\hbar):=[E-\hbar \omega,
E+\hbar \omega]$, which shall not contain any critical values of $H_0$,
there exists a sub-sequence $\{\psi_{k_\nu}\, |\, E_{k_\nu}\in I(E,\hbar)\}$ 
of density one, i.e.\
\begin{equation}
\label{densone}
 \lim_{\hbar \to 0} \frac{\# \{ \nu\, | \, E_{k_\nu} \in I(E,\hbar) \}}
 {\# \{ k \, | \, E_k \in I(E,\hbar) \}} =1 \ ,
\end{equation}
such that for every quantum observable $\hat B$ with hermitian symbol 
$B(\vecp,\vecx)$ and principal symbol $B_0(\vecp,\vecx)$ the expectation 
values taken in the eigenstates $\{ \psi_{k_\nu} \}_{\nu \in \nz}$ of 
$\hat H_P$ fulfill
\begin{equation}
\label{QE}
 \lim_{\nu\to \infty} \langle \psi_{k_\nu}, \hat B \psi_{k_\nu} \rangle 
 = \frac{1}{2} \mtr \int_{\Om_E} B_0(\vecp,\vecx) \ \ud\mu_E(\vecp,\vecx)
 =: \frac{1}{2} \mtr \mu_E(B_0) \ .
\end{equation}
Moreover, the sub-sequence $\{ \psi_{k_\nu} \}_{\nu\in \nz}$ can be chosen 
independent of the observable $\hat B$. Here we remark that since the
number $N_I$ of eigenvalues $E_k\in I(E,\hbar)$ grows like
\begin{equation}
\label{NIgrowth}
N_I \sim 2\hbar\om\,\frac{(2j+1) \vol\Om_E}{(2\pi\hbar)^d}
\end{equation}
(see \cite{BolKep99b,BolGla00}) in the semiclassical limit, both the numerator 
and the denominator in (\ref{densone}) become infinite as $\hbar\to 0$. Thus, 
the limit $\nu\to\infty$ in (\ref{QE}) requires the semiclassical limit.

We will now derive the analogous result for Pauli Hamiltonians with 
arbitrary spin under the condition that $Y_{\rm cl}^t$, i.e. the purely
classical skew product (\ref{skew2}), is ergodic on $\Om_E \times \Stwo$ 
with respect to the invariant measure 
$\mu_{\rm cl} := \mu_E \times \mu_{\Stwo}$. This means that for every 
integrable function $f\in L^1(\Om_E\times\Stwo, \ud\mu_{\rm cl})$ the relation 
\begin{equation} 
\label{S2ergodic}
 \lim_{T \to \infty} \frac{1}{T} \int_0^T 
 f(Y_{\rm cl}^t(\vecxi,\vecy,\vecm)) \, \ud t = 
 \int_{\Om_E \times \SU(2)} f(\vecp,\vecx,\vecn) 
 \ \ud \mu_{\rm cl}(\vecp,\vecx,\vecn)
\end{equation}
holds for $\mu_{\rm cl}$-almost all initial conditions $(\vecxi,\vecy,\vecm) 
\in \Om_E \times \Stwo$. We remark that the ergodicity (\ref{Yergod}) of 
$Y^t$, which was required for quantum ergodicity in \cite{BolGla00}, implies 
the relation (\ref{S2ergodic}), i.e. ergodicity of $Y_{\rm cl}^t$, see 
appendix \ref{ergodicities}. Also notice that using property (iii) on page 
\pageref{symbProp}, the right-hand side of (\ref{QE}) can be rewritten as 
($j=\frac{1}{2}$) 
\begin{equation}
  \frac{1}{2} \mtr \mu_E(B_0) =  
  \int_{\Om_E\times\Stwo} b_0(\vecp,\vecx,\vecn) \ 
  \ud \mu_{\rm cl}(\vecp,\vecx,\vecn) =:\mu_{\rm cl}(b_0)\ .
\end{equation}

In order to prove quantum ergodicity for arbitrary spin $j$ along the 
lines of \cite{Zel96,ZelZwo96,BolGla00}, we now analyse the quantity  
\begin{equation}
\label{S2}
 S_2(E,\hbar):= \frac{1}{N_I} \sum_{E_k \in I(E,\hbar)} \left| 
 \langle \psi_k, \hat B \psi_k \rangle - \mu_{\rm cl}(b_0) \right|^2 \ ,
\end{equation}
which is the variance of the expectation values of the quantum observable 
$\hat B$ about the classical mean of its scalar principal symbol $b_0$. 
Introducing the bounded and self-adjoint auxiliary operator
\begin{equation}
\label{auxop}
 \hat B_T:= \frac{1}{T} \int_0^T \ue^{\frac{\ui}{\hbar} \hat H_P t}\, \hat B 
 \, \ue^{- \frac{\ui}{\hbar} \hat H_P t} \ \ud t -  
 \mu_{\rm cl}(b_0)\, \eins_{2j+1} \ ,
\end{equation}
the variance (\ref{S2}) also reads
\begin{equation} 
\label{S2mod}
 S_2(E,\hbar)= \frac{1}{N_I} \sum_{E_k \in I(E,\hbar)} \left| \langle
 \psi_k, \hat B_T \psi_k \rangle \right|^2 
 \leq \frac{1}{N_I} \sum_{E_k \in I(E,\hbar)} \langle
 \psi_k, \hat B_T^2 \psi_k \rangle \ ,
\end{equation}
where the estimate on the right-hand side follows from the Cauchy-Schwarz 
inequality. Quantum ergodicity is now implied, if this upper bound vanishes 
in the semiclassical limit. In order to show that this indeed happens,
we employ the Szeg\"o limit formula 
\begin{equation} 
\label{szegoe}
 \lim_{\hbar \to 0} \frac{1}{N_I} \sum_{E_k \in I(E,\hbar)} 
 \langle \psi_k, \hat B \psi_k \rangle = \frac{1}{2j+1} \, \mtr \mu_E(B_0)
 = \mu_{\rm cl}(b_0) 
\end{equation}
for an observable $\hat B$, which has been established in \cite{BolGla00}. 
This formula is valid under certain technical assumptions on the symbol of 
the Hamiltonian $\hat H_P$ (see \cite{BolGla00} for details), and if the 
periodic orbits of the translational flow $\Phi^t$ with non-trivial periods 
form a set of Liouville measure zero in $\Om_E$. Since $Y^t_{\rm cl}$ shall 
be ergodic, which includes the ergodicity of the translational motion on the
energy shell $\Om_E$, this condition is clearly fulfilled and 
we can apply (\ref{szegoe}) to (\ref{S2mod}), yielding
\begin{equation}
\label{semlimS2}
 \lim_{\hbar \to 0} S_2(E,\hbar) \leq \frac{1}{2j+1} \mtr \mu_E 
 \bigl( (B_{T,0})^2\bigr)
  = \mu_{\rm cl} \bigl( (b_{T,0})^2 \bigr)\ ,
\end{equation}
where on the rigth-hand side use has been made of the relation 
(\ref{products}). Here $B_{T,0}(\vecp,\vecx)$ and $b_{T,0}(\vecp,\vecx,\vecn)$
denote the matrix valued and scalar principal symbol, respectively, of the 
auxiliary operator (\ref{auxop})

In order to conclude the proof consider the scalar principal symbol 
$b_{T,0}$ of $\hat{B}_T$, which according to (\ref{auxop}) and 
(\ref{scalarEgorov2}) reads
\begin{equation}
\label{bT0}
 b_{T,0}(\vecp,\vecx, \vecn) = 
 \frac{1}{T} \int_0^T b_0(Y^t_{\rm cl}(\vecp,\vecx, \vecn)) \ \ud t 
 - \mu_{\rm cl}(b_0) \ .
\end{equation}
An application of the ergodicity (\ref{S2ergodic}) of $Y_{\rm cl}^t$ to the 
principal symbol $b_{0}(\vecp,\vecx,\vecn)$ of the observable $\hat{B}$
now reveals that $b_{T,0}
(\vecp,\vecx,\vecn)$ vanishes as $T\to\infty$ for $\mu_{\rm cl}$-almost all 
$(\vecp,\vecx,\vecn)\in\Om_E \times\Stwo$. Since therefore the right-hand side
of (\ref{semlimS2}) vanishes, this finally implies that  
\begin{equation}        
  \lim_{\hbar \to 0} S_2(E,\hbar)=0 \ .
\end{equation}
According to a general construction in proofs of quantum ergodicity 
\cite{CdV85,Zel87}, the vanishing of $S_2(E,\hbar)$ in the semiclassical 
limit implies the existence of a density-one sub-sequence 
$\{ \psi_{k_\nu} \}_{\nu\in\nz} \subset \{ \psi_k \}_{k \in \nz }$ along which 
one has 
\begin{equation}
\label{QES2}
 \lim_{\nu\to \infty} \langle \psi_{k_\nu}, \hat B \psi_{k_\nu} \rangle 
 = \mu_{\rm cl}(b_0) \ .
\end{equation}
Finally, a diagonal construction (see again \cite{CdV85,Zel87}) yields a 
sub-sequence, still of density one, that is independent of the observable 
$\hat B$ such that (\ref{QES2}) holds along this sub-sequence for all
observables of the type considered here. 

Our conclusion therefore is that quantum ergodicity for Pauli Hamiltonians 
with arbitrary spin $j$ holds under the condition that the classical skew
product $Y_{\rm cl}^t$ of translational and spin dynamics is ergodic on 
$\Omega_E\times\Stwo$. We remark that ergodicity of $Y_{\rm cl}^t$ implies 
ergodicity of the translational motion $\Phi^t$ on the energy shell 
$\Omega_E$. However, ergodicity of $\Phi^t$ alone is not a sufficient 
condition for quantum ergodicity; see \cite{BolGla00} for a 
counter-example.

For the purpose of an interpretation of the quantum ergodicity relation
(\ref{QES2}) for Pauli Hamiltonians with arbitrary spin, we now discuss its 
effect in terms of Wigner transforms of spinors $\psi\in\cS'(\rz^d)
\otimes\kz^{2j+1}$. Matrix valued Wigner transforms are defined as
\begin{equation}
\label{wignertf}
 W[\psi](\vecp,\vecx):= \int_{\rz^d} \ue^{-\frac{\ui}{\hbar} \vecp \vecy} \,
 \overline{\psi \left( \vecx - \tfrac{1}{2}\vecy \right)} \otimes 
 \psi\left( \vecx + \tfrac{1}{2}\vecy \right) \ \ud y \ ,
\end{equation}
so that quantum expectation values of an observable $\hat B$ with symbol
$B(\vecp,\vecx)$ in states given by $\psi \in L^2(\rz^d) \otimes \kz^{2j+1}$ 
read
\begin{equation}
\label{expvalwig}
 \langle \psi, \hat B \psi \rangle = \frac{1}{(2 \pi \hbar)^d} \int_{\rz^d} 
 \int_{\rz^d} \mtr \left( W[\psi](\vecp,\vecx) \, B(\vecp,\vecx) \right) \
 \ud p \, \ud x \ .
\end{equation}
After employing the relation (\ref{products}) we obtain 
\begin{equation}
\label{EVwig_sc}
 \langle \psi, \hat B \psi \rangle = \frac{2j+1}{(2\pi\hbar)^d} \int_{\rz^d} 
 \int_{\rz^d} \int_{\Stwo} w[\psi](\vecp,\vecx,\vecn)\, b(\vecp,\vecx,\vecn) 
 \ \ud p \, \ud x \, \ud\mu_{\Stwo}(\vecn) \ ,
\end{equation}
where
\begin{equation}
\label{scWigner}
 w[\psi](\vecp,\vecx,\vecn) := 
 \mtr \left( W[\psi](\vecp,\vecx) \, \hat{\Delta}_j(\vecn) \right)  
\end{equation} 
is introduced as a scalar Wigner transform of $\psi\in\cS'(\rz^d)\otimes
\kz^{2j+1}$. In this context, quantum ergodicity (\ref{QES2}) implies that 
along a density-one sub-sequence the scalar Wigner transforms 
$\{ w[\psi_{k_\nu}] \}_{\nu \in \nz}$ of eigenspinors of the quantum 
Hamiltonian 
become equidistributed on both the (translational) energy shell $\Om_E$ and 
on the classical phase space $\Stwo$ of spin, i.e.
\begin{equation}
\label{wignerConv}
  \lim_{\nu \to \infty} \frac{2j+1}{(2 \pi \hbar)^d} \, 
  w[\psi_{k_\nu}](\vecp,\vecx,\vecn) 
  = \frac{1}{\vol \Omega_E} \, \delta(H_0(\vecp,\vecx) -E) \, ,
\end{equation}
where the convergence has to be understood in the sense of distributions,
i.e. after integration with a suitable function on 
$\rz^d\times\rz^d\times\Stwo$ as in (\ref{EVwig_sc}).


\subsection*{Acknowledgment}

We would like to thank the Deutsche Forschungsgemeinschaft for financial 
support under contracts no. Ste 241/10-1 and no. Ste 241/15-1.

\begin{appendix}
\section{Explicit form of the kernel $\hat{\Delta}_j(\vecn)$}
\label{kernels}
In this appendix we comment on some properties of the particular 
representation for the kernel $\hat{\Delta}_j(\vecn)$ that was chosen in 
\cite{VarGra89} and give explicit examples for $j=\frac{1}{2}$ and $j=1$.

Expressing $\vecn \in \Stwo \subset \rz^3$ in spherical coordinates, 
$\vecn = (\sin\theta\cos\vp,\sin\theta\sin\vp,\cos\theta)^T$, we denote
the usual spherical harmonics by $Y_{lm}(\vecn)$. Then the kernel 
$\hat{\Delta}_j(\vecn)$ can be expanded in the basis 
$\{Y_{lm}(\vecn) \, | \, l \in \nz_0,|m|\leq l\}$,
\begin{equation}
  \hat{\Delta}_j(\vecn) 
  = \sum_{l=0}^{2j} \sum_{m=-l}^l 
    \sqrt{\frac{4\pi}{2j+1}} \, C_{lm}^{j\dagger} \, Y_{lm}(\vecn) \ ,
\end{equation}  
where for given $j$ the $C_{lm}^j$ are $(2j+1)\times(2j+1)$ matrices, 
see \cite{VarGra89} for details. Notice that this expansion terminates at 
$l=2j$.

Thus, if one expands a general classical observable, given by some function 
$b(\vecn)\in L^2(\Stwo)$, in spherical harmonics and uses the spin-Weyl
correspondence (\ref{spinWeyl}) in order to associate to it a spin-$j$ quantum 
observable $\hat{b}$, then its Weyl symbol (\ref{spinsymbol}) will in general 
not coincide with the original $b(\vecn)$, but instead its expansion in 
spherical harmonics will be truncated at $l=2j$. However, since the fundamental
theory is quantum mechanics, one should adopt the reverse point of view: 
Starting from a spin-$j$ quantum observable $\hat{b}$ one can assign to it 
its Weyl symbol $b(\vecn)$, given by (\ref{spinsymbol}), as a classical 
observable. Then by (\ref{spinWeyl}) this correspondence is one-to-one.

Therefore, in a classical description of a spin $j$ a general observable 
$b(\vecn)\in L^2(\Stwo)$ has an expansion in spherical harmonics 
$Y_{lm}(\vecn)$ which terminates at $l=2j$. Conversely, if the aim were a 
complete quantum description of a top whose classical dynamics is given 
by the Euler equations (instead of a 
classical description of, say, an electron with fixed spin $\frac{1}{2}$), 
one would have to perform the semiclassical limit $\hbar \to 0$ simultaneously
with the limit of large spin, $j\to\infty$, such that $j\hbar = O(1)$,
cf., e.g., \cite{Haa00}.

As an explicit example we now briefly discuss some properties of 
$\hat{\Delta}_j(\vecn)$ in the representation chosen in \cite{VarGra89}. 
For spin $j=\frac{1}{2}$ we obtain 
\begin{equation}
  \hat{\Delta}_{1/2}(\vecn) 
  = \sqrt{\pi} \left( \begin{matrix} 
      Y_{00}(\vecn) + Y_{10}(\vecn) & - \sqrt{2} \ \overline{Y_{11}(\vecn)} \\
      - \sqrt{2} \ Y_{11}(\vecn) & Y_{00}(\vecn) - Y_{10}(\vecn)
      \end{matrix} \right)
  = \tfrac{1}{2} \eins_2 + \sqrt{\tfrac{3}{4}} \, \vecn \cdot \vecsig \, .
\end{equation}
Since every observable $\hat b$ of a spin 1/2 is a hermitian $2\times 2$
matrix, which can be represented as $\hat b=b_0\,\eins_2+\vecb\cdot\vecsig$
with $b_0\in\rz$ and $\vecb\in\rz^3$, the associated classical observable
reads
\begin{equation}
b(\vecn) = \mtr\bigl( \hat b\,\hat{\Delta}_{1/2}(\vecn) \bigr) = 
b_0 + \sqrt{3}\,\vecb\cdot\vecn 
=b_0 + 2 \sqrt{j(j+1)}\,\vecb\cdot\vecn \ .
\end{equation}
In particular, the spin operator $\hat{\vecS}=\frac{\hbar}{2}\vecsig$ is
mapped to the symbol $\vecs = \hbar \sqrt{3/4}\,\vecn$. 
\kommentar{
For $j=1$ the kernel reads
\begin{equation}
\begin{split}   
  &\hat{\Delta}_{1}(\vecn) \\ 
  &= \left( \begin{matrix} 
  \frac{1}{3} \left( 2\sqrt{\pi} Y_{00}{} 
  + \sqrt{2\pi} (-\sqrt{3}Y_{10}{}+Y_{20}{}) \right) &
  \sqrt{\frac{2\pi}{3}} \left( -Y_{11}{}+Y_{21}{} \right) &
  2\sqrt{\frac{\pi}{3}} Y_{22}{} 
  \\
  \sqrt{\frac{2\pi}{3}} \left( 
  -\overline{Y_{11}{}}+\overline{Y_{21}{}} \right) &
  \frac{2}{3} \left( \sqrt{\pi} Y_{00} - \sqrt{2\pi} Y_{20} \right) &
  -\sqrt{\frac{2\pi}{3}} \left( Y_{11}{} + Y_{21}{} \right)
  \\
  2\sqrt{\frac{\pi}{3}} \overline{Y_{22}{}} &
  -\sqrt{\frac{2\pi}{3}} \left( 
  \overline{Y_{11}{}} + \overline{Y_{21}{}} \right) &
  \frac{1}{3} \left( 2\sqrt{\pi} Y_{00}{} 
  + \sqrt{2\pi} (\sqrt{3}Y_{10}{}+Y_{20}{}) \right)
  \end{matrix} \right)^T
\end{split}     
\end{equation}
}
This last statement generalises to arbitrary spin, i.e. 
$\hat{\vecS}=\frac{\hbar}{2} \ud \pi_j(\vecsig)$ is always mapped to 
$\vecs = \hbar \sqrt{j(j+1)}\,\vecn$. E.g., for $j=1$ the inverse 
transformation (\ref{spinWeyl}) reads
\begin{equation}
\begin{split}   
  \hat{\vecS} 
  &= 3 \int_{\Stwo} \vecs \, \hat{\Delta}_{1}(\vecn) \, \ud\mu_{\Stwo}(\vecn) 
  \\
  &= \frac{\hbar}{2} \left(  
  \left( \begin{matrix} 0 & \sqrt{2} & 0 \\
                             \sqrt{2} & 0 & \sqrt{2} \\
                             0 & \sqrt{2} & 0 \end{matrix} \right) ,
  \left( \begin{matrix} 0 & -\ui\sqrt{2} & 0 \\
                        \ui\sqrt{2} & 0 & -\ui\sqrt{2} \\
                        0 & \ui\sqrt{2} & 0 \end{matrix} \right) ,
  \left( \begin{matrix} 2 & 0 &  0 \\
                             0 & 0 &  0 \\
                             0 & 0 & -2 \end{matrix} \right)
  \right)^T ,
\end{split}     
\end{equation}
where now the kernel $\hat{\Delta}_1(\vecn)$ can be expressed as
\begin{equation}
  \hat{\Delta}_1(\vecn) = \frac{1-\sqrt{10}}{3} \, \eins_3 
  + \frac{\sqrt{2}}{4} \, (\ud\pi_1(\vecsig) \cdot \vecn) 
  + \frac{\sqrt{10}}{8} \, \left(\ud\pi_1(\vecsig) \cdot \vecn \right)^2 \, .
\end{equation}

\section{Ergodicity of $Y^t$ and $Y_{\rm cl}^t$}
\label{ergodicities}

In this appendix we discuss the relation between the two skew products 
$Y^t$ (\ref{skew1}) and $Y_{\rm cl}^t$ (\ref{skew2}) and their 
respective ergodic properties. 

To this end we need a mapping of quantum spin precession on $\SU(2)$ 
to classical spin precession on $\Stwo$. This is conveniently given by the 
Hopf map, one of whose possible realisations reads 
\begin{equation}
\begin{split}  
  \pi_H: \ \SU(2) \quad & \longrightarrow \quad \Stwo \\
           g     \qquad & \longmapsto u^\dagger g^\dagger \vecsig g u \ ,
\end{split}     
\end{equation}  
where $u \in \kz^2$ with $\|u\|_{\kz^2}=1$ is arbitrary but fixed. We now want
to relate the spin part $g\mapsto d(\vecp,\vecx,t)g$ of the skew product flow
$Y^t$ to the spin part $\vecn\mapsto\vp(d(\vecp,\vecx,t))\vecn$ of 
$Y_{\rm cl}^t$. The definition of the universal covering map 
$\vp: \, \SU(2) \to \SO(3)$ implies that
\begin{equation}
\label{evolHopf}
  \pi_H(d(\vecp,\vecx,t)g) = \varphi(d(\vecp,\vecx,t)) \, \pi_H(g) \ ,
\end{equation}
so that the projection of the (quantum mechanical) time evolution in $\SU(2)$
to $\Stwo$ fulfills the classical equation of spin precession 
(\ref{clprecess}). This is a manifestation of the Ehrenfest theorem, since
the Hopf map (\ref{evolHopf}) yields the expectation value of (normalised)
quantum spin in the state $gu\in\kz^2$, see \cite{BolKep99a}. 

Therefore, defining $\tilde{\pi}_H := \id_{\rz^d\times\rz^d} \otimes \, \pi_H: 
 \, \rz^d\times\rz^d\times\SU(2) \to \rz^d\times\rz^d\times\Stwo$, i.e.
$\tilde{\pi}_H (\vecp,\vecx,g) = (\vecp,\vecx,\pi_H(g))$, we
obtain the following commuting diagram:
\begin{equation}\nonumber       
\begin{CD}
  (\vecp,\vecx,g) 
    @>{\hspace{2.5cm} \displaystyle Y^t \hspace{2.5cm}}>> 
  (\Phi^t(\vecp,\vecx),d(\vecp,\vecx,t)g) \\
    @VV{\begin{matrix} \ \\ \tilde{\pi}_H \\ \ \end{matrix}}V  
    @VV{\begin{matrix} \ \\ \tilde{\pi}_H \\ \ \end{matrix}}V \\
  (\vecp,\vecx,\vecn)
    @>{\hspace{2.5cm} \displaystyle Y_{\rm cl}^t \hspace{2.5cm}}>>
  (\Phi^t(\vecp,\vecx),\varphi(d(\vecp,\vecx,t))\vecn) 
\end{CD} 
\end{equation}  
Furthermore, to every (integrable) function 
$f \in L^1(\Om_E \times \Stwo, \ud\mu_{\rm cl})$ one can associate a function 
$F \in L^1(\Om_E \times \SU(2),\ud\mu)$ by defining $F:=f\circ\tilde{\pi}_H$,
i.e. $F(\vecp,\vecx,g)=f(\vecp,\vecx,\pi_H(g))$. Provided now that $Y^t$ 
is ergodic on $\Om_E\times\SU(2)$ with respect to $\mu=\mu_E\times\mu_H$, 
one observes that
\begin{equation} 
\label{SU2toS2}
\begin{split} 
 \lim_{T \to \infty} \frac{1}{T} \int_0^T 
 f\bigl(Y^t_{\rm cl}(\vecxi,\vecy,\pi_H(h))\bigr) \ \ud t 
 & = \lim_{T \to \infty} \frac{1}{T} \int_0^T 
     F(Y^t(\vecxi,\vecy,h)) \ \ud t \\
 & = \int_{\Om_E \times \SU(2)} F(\vecp,\vecx,g) \ \ud \mu(\vecp,\vecx,g) \\ 
 & = \int_{\Om_E \times \Stwo} f(\vecp,\vecx,\vecn) \ 
     \ud\mu_{\rm cl}(\vecp,\vecx,\vecn) 
\end{split} 
\end{equation}
for $\mu_E$-alomst all $(\vecxi,\vecy)\in\Om_E$ and $\mu_H$-almost
all $h\in\SU(2)$. Since this implies the validity of (\ref{SU2toS2}) for
$\mu_{\Stwo}$-almost $\vecm=\pi_H(h)\in\Stwo$, the ergodicity of $Y^t$ 
implies ergodicity of $Y^t_{\rm cl}$ on $\Om_E\times\Stwo$ with respect to 
$\mu_{\rm cl} = \mu_E \times \mu_\Stwo$.

\end{appendix}

\bibliographystyle{my_unsrt}                   
\bibliography{literatur}                       

\end{document}